\documentstyle[12pt]{article}

\newcommand{\be}{\begin{equation}}
\newcommand{\ee}{\end{equation}}
\newcommand{\bea}{\begin{eqnarray}}
\newcommand{\eea}{\end{eqnarray}}

\newcommand{\nn}{\nonumber}

\newcommand{\bi}{\bar i}

\newcommand{\RR}{\rangle}
\newcommand{\LL}{\langle}

\makeatletter
\newdimen\normalarrayskip              
\newdimen\minarrayskip                 
\normalarrayskip\baselineskip
\minarrayskip\jot
\newif\ifold             \oldtrue            

\makeatother
\newlength{\extraspace}
\newlength{\extraspaces}
\input{tcilatex}

\begin{document}

\addtolength{\baselineskip}{.4mm}

\thispagestyle{empty}

\begin{flushright}
\baselineskip=12pt
$         $\\
\hfill{  }\\ 
\end{flushright}
\vspace{.5cm}

\begin{center}
\baselineskip=24pt

{\LARGE {Entanglement Swapping Chains for General Pure States}}\\[15mm]

\baselineskip=12pt

{Lucien Hardy and David D. Song} \\[%
8mm]
{ Centre for Quantum Computation\\[0pt]
Clarendon Laboratory, Department of Physics \\ 
University of Oxford, Parks Road, Oxford OX1 3PU, U.K.} \vspace{2cm}

{\bf Abstract}

\begin{minipage}{15cm}
\baselineskip-12pt

We consider entanglement swapping schemes with general (rather than
maximally) entangled bipartite states of arbitary dimension shared pairwise
between three or more parties in a chain.  The intermediate parties
perform generalised Bell measurements with the result that the two end
parties end up sharing a entangled state which can be converted
into maximally entangled states.  We obtain
an expression for the average amount of maximal entanglement
concentrated in such a scheme and show that in a certain reasonably
broad class of cases this scheme is provably optimal and that, in these cases,
the amount of entanglement concentrated between the two ends is equal to
that which could be concentrated from the weakest link in the chain.

\end{minipage}
\end{center}

\vspace{.2cm}

\section{Introduction}
There are many applications of quantum entanglement including quantum cryptography \cite{brassard1,ekert,brassard2},
 teleportation \cite{bennett1} and quantum communication \cite{cleve}. 
These applications often require maximally entangled states (that is
states of the form
$|\varphi_m\RR=\frac{1}{\sqrt m} \sum_{i=1}^m |i\RR_A|i\RR_B$).
However bipartite entanglement shared by two parties may well not be maximal.
A number of schemes for obtaining maximally entangled states
from non-maximal ones have been
investigated both in cases where the starting state is pure
(the process is then called entanglement concentration) and where it is
mixed (the process is then called purification) \cite{smolin1,smolin2,gisin,deutsch}. More generally, schemes
have also been considered for manipulating general entangled states to
other general entangled states by local operations and classical communication (LOCC)
 \cite{nielsen,catalyst}.
In this paper we
will be concerned only with concentrating the entanglement of pure states.
These schemes assume that the distant parties sharing the entangled
state are only allowed to perform LOCC.
 In \cite{bennett2}, it was  shown that $N$ copies of a non-maximal two particle entangled state
can be converted to maximally entangled states by LOCC in the asymptotic limit $N
\rightarrow \infty$.
Subsequently, \cite{lo,hardy,jonathan}, various people showed
how to convert a single copy of a general entangled state to a
distribution of known maximally entangled states.
In such a scheme we consider the process where, under LOCC, we can
convert a general pure state $|\psi_m\RR$
to a distribution of maximally entangled states $
|\varphi_m\RR=\frac{1}{\sqrt m} \sum_{i=1}^m |i\RR_A|i\RR_B$ where
$|\varphi_m\RR$ occurs with probability, $p_m$.  The state $|\varphi_m\RR$ is given weighting $\log_2 m$
since it corresponds to $\log_2 m$ copies of $|\varphi_2\RR$ \cite{bennett2,lo}. The average maximal
entanglement produced is then $E=\sum_m p_m \log_2 m$.
In \cite{hardy,jonathan}, it was shown that for one copy of a general
bipartite pure state
in Schmidt form, $|\psi\RR= \sum_{i=0}^{m-1} \sqrt{\lambda_i}|i\RR_A
|i\RR_B$ with $\lambda_i\geq\lambda_{i+1}$,
the maximum average entanglement that can be concentrated in the
form of maximally entangled state is given by
\be
E^{\rm max} = \sum_{i=1}^m (\lambda_{i-1} - \lambda_i ) i\log_2 i
\label{Emax}\ee
(where $\lambda_m=0$).
Hence, for a pure entangled state shared between two parties, we have a
full understanding of how to concentrate entanglement.

In this paper we wish to consider more general schemes in which
entanglement may be concentrated.  In particular, we will consider
entanglement swapping. Entanglement swapping was introduced in \cite{zukowski}
(and also in \cite{bennett1}). The idea is that Alice shares an
entangled state with Bob and Bob shares another entangled state with
Charlie.  Bob performs a Bell measurement on his two particles and the
result is that Alice and Charlie end up sharing an entangled state.
Initially, only the case where Alice and Bob and likewise Bob and Charlie
shared maximally entangled states was considered (or in the case of
\cite{bennett1} at lease one of the states was maximally entangled
allowing teleportation).  Later, Bose, Vedral and Knight considered
the case \cite{bose} where both states are of the form $\alpha|00\RR+\beta|11\RR$.
They found, perhaps rather surprisingly, that the amount of entanglement
concentrated corresponds to the optimal amount which could be
concentrated were the state $\alpha|00\RR+\beta|11\RR$ shared directly
between Alice and Charlie (namely $2|\beta|^2$ where
$|\beta|\leq|\alpha|$).  Very recently, Shi, Jiang and Guo \cite{recent} considered the
case where Alice and Bob share the state $\alpha|00\RR+\beta|11\RR$ and
Bob and Charlie share the state $\alpha'|00\RR+\beta'|11\RR$.  This time
it turns out that the amount of entanglement concentrated between Alice
and Charlie corresponds to that which could be concentrated from
the less entangled of these two states were it shared directly between
Alice and Charlie. In this case the entanglement concentrated
corresponds to that of the weakest link.  
Entanglement swapping chains have also been considered with impure states \cite{zoller}. 
The objective there was to exchange pure entanglement over long distance through noisy channels.

In this paper we will generalise this situation in two respects.  We
will consider the case where the states shared are of $m\times m$
dimensions and we will consider chains consisting of three or more
parties with generalised Bell measurements at all the intermediate
locations.  We will obtain a result for the average amount of entanglement that
can concentrated (in the form of maximally entangled states) between the
two parties by this method.  We find that for a particular rather broad
class of
cases (but not in all cases) the average amount of entanglement
concentrated corresponds to weakest link in the chain.  Namely that in
these cases, the average entanglement concentrated is equal to that which could
be concentrated were that state which has lowest $E^{\rm max}$ (as given
by equation (1)) shared between the two end parties.  Therefore, in
these cases the protocol is optimal.  However, for other cases we do not
know that the protocol is optimal - it is possible that measurements
other than generalised Bell measurements could yield a higher average.

In section 2 we review the case where three parties share two $2\times
2$ dimensional states as first considered by \cite{recent}.
In section 3 we generalise this to $m\times m$ dimensional states and in
section 4  we generalise further to the case where a number of parties
 arranged in a chain share entangled states pairwise along the chain.
Finally, in section 5 we consider the case where GHZ states
are concentrated from two $2\times 2$ states shared between three
parties.

\section{ $2 \times 2$ case}
Suppose Alice and Bob share a general bipartite entangled state
$|\psi^{(1)}\RR$ and Bob and Charlie share another entangled state
$|\psi^{(2)}\RR$.  We can always write these states in Schmidt form:
\bea
|\psi^{(1)}\RR &=& \sqrt{\lambda_0^{(1)}}|00\RR_{12} + \sqrt{\lambda_1^{(1)}}|11\RR_{12} \label{221} \\
|\psi^{(2)}\RR &=& \sqrt{\lambda_0^{(2)}}|00\RR_{34} + \sqrt{\lambda_1^{(2)}}|11\RR_{34}\label{222}
\eea
where the $\lambda$'s are real and non-negative and satisfy
$\sum_{j=0}^1 \lambda_j^{(i)} =1$ for $i=1,2$ (normalisation) and are
taken to be ordered so that $\lambda_0^{(n)} \geq \lambda_1^{(n)} $ for
$n=1,2$. Also, for simplicity, we assume that $\lambda_0^{(1)} \geq
\lambda_0^{(2)}$ (this means that the state Alice and Bob share has less
than or equal entanglement to the state that Bob and Charlie share).
As shown in Figure 1, particles 1 and 4 are spatially separated. 
%
%

\setlength{\unitlength}{1mm}

\begin{picture}(120,40)

\put(15,20){\line(1,0){15}}
\put(15,20){\circle*{2}}
\put(30,20){\circle*{2}}
\put(35,20){\line(1,0){15}}
\put(35,20){\circle*{2}}
\put(50,20){\circle*{2}}
\put(14,14){1}
\put(29,14){2}
\put(34,14){3}
\put(49,14){4}

\put(5,3){\framebox(55,30)}

\put(67,18){\vector(1,0){10}}

\put(95,20){\line(1,0){35}}
\put(95,20){\circle*{2}}
\put(94,14){1}
\put(129,14){4}
\put(130,20){\circle*{2}}
\put(85,3){\framebox(55,30)}

\end{picture}

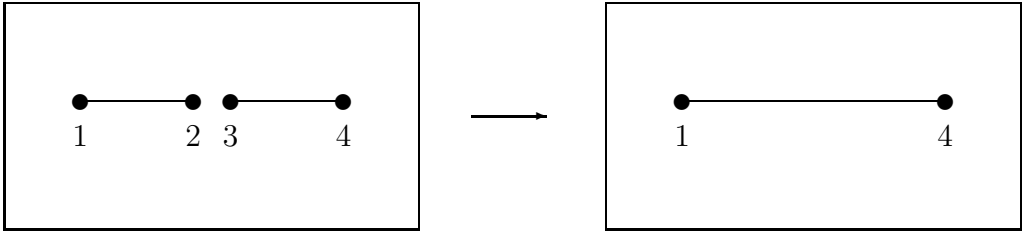
Figure 1: The swapping of entanglement of two bipartite states is shown.  A Bell measurement is made on particles 2
 and 3.
\vspace{1cm}
\newline
Bob performs Bell measurement on his particles
2 and 3 onto the basis,
 $ \frac{1}{\sqrt 2} (|00\RR_{23} \pm |11\RR_{23} )$ and
$ \frac{1}{\sqrt 2} (|10\RR_{23} \pm |10\RR_{23} ) $.
Then the particles 1 and 4 for Alice and Charlie are projected onto
\bea
|\psi\RR_{14}^{\pm}&=   & P_1^{-1/2} \left( \sqrt{\frac{\lambda_0^{(1)} \lambda_0^{(2)}}{2}} |00\RR_{14}  \pm \sqrt{\frac{\lambda_1^{(1)} \lambda_1^{(2)}}{2}} |11\RR_{14} \right) \label{4states1} \\
|\phi\RR_{14}^{\pm} & =   & P_2^{-1/2} \left( \sqrt{\frac{\lambda_0^{(1)} \lambda_1^{(2)}}{2}} |00\RR_{14}  \pm \sqrt{\frac{\lambda_0^{(2)} \lambda_1^{(1)}}{2}} |11\RR_{14} \right)
\label{4states2}\eea
where $P_1$ and $P_2$ are probabilities for getting $|\psi\RR_{14}^{\pm}$ and $|\phi\RR_{14}^{\pm}$ respectively which are 
\bea
P_1 &=& \frac{\lambda_0^{(1)} \lambda_0^{(2)}}{2} + \frac{\lambda_1^{(1)} \lambda_1^{(2)}}{2} \nonumber \\
P_2  &=& \frac{\lambda_0^{(1)} \lambda_1^{(2)}}{2} + \frac{\lambda_0^{(2)} \lambda_1^{(1)}}{2}
\eea
It follows from (\ref{Emax}) that for a general bipartite two-level state, 
the maximum average entanglement concentrated in the form of
maximally entangled states is
twice the square of the lower coefficient, among two of them. In the
$2\times 2$ case this is also equal to the maximum probability of
obtaining a $|\varphi_2\RR$ state.
 Since we assumed $\lambda_0^{(1)} \geq \lambda_0^{(2)} \geq \lambda_1^{(2)} \geq \lambda_1^{(1)}$, it follows  
$\lambda_0^{(1)}\lambda_0^{(2)} \geq \lambda_1^{(1)}\lambda_1^{(2)}$ and $\lambda_0^{(1)}\lambda_1^{(2)} \geq \lambda_0^{(2)}\lambda_1^{(2)}$.
  Therefore the probability of getting maximally entangled state between 1 and 4 with four states in (\ref{4states1}) and (\ref{4states2}) is 
\be
P_1 \left(  \frac{2 \lambda_1^{(1)} \lambda_1^{(2)}}{P_1} \right) + P_2 \left( \frac{2 \lambda_1^{(1)}\lambda_0^{(2)}}{P_2} \right)  = 2\lambda_1^{(1)}\lambda_1^{(2)} + 2\lambda_1^{(1)} \lambda_0^{(2)} = 2\lambda_1^{(1)} 
\label{2lambda}\ee
This result is optimal because $2\lambda_1^{(1)}$ actually corresponds
to the maximum probability of Alice and Bob being able
to share a maximally entangled state. We see here that Alice and Charlie are able to share as
much maximal entanglement as can be extracted from the weakest link in the chain.

The above procedure can be illustrated with the method of area diagrams introduced in \cite{hardy}.
After cancellation of probability with the normalisation constant and taking the $\pm$ state together, 
(\ref{4states1}) and (\ref{4states2}) can be put into area diagrams $(i)$ and $(ii)$ of (a) in Figure 2. 
 Each box has unit 
width, so the height of the box is equal to its area.
$E^{\rm max}$ can be calculated from column $(b)$ of $(i)$ and $(ii)$ 
by multiplying the area of that box of width $i$ by $\log_2
i$ then sum over all $i$'s (i.e. $i=1,2$).
$E^{\rm max}$ can also be calculated by first adding the boxes of $(i)$
and $(ii)$ therefore get $(iii)$ of (a), then
calculate from the boxes in (b) of $(iii)$.  Once $(i)$ and $(ii)$ are
added, the terms $\lambda_0^{(2)}$
and $\lambda_1^{(2)}$ add up to 1 then it is easy to see that $E^{\rm
max}$ is equal to $E^{\rm max}$ for $|\psi^{(1)}\RR$ only,  which is $2\lambda_1^{(1)}$.
This graphical method is very useful in the general $m\times m$ case.

%
%

\begin{picture}(120,175)

\put(10,152){$(i)$}
\put(35,140){\line(0,1){27}}
\put(35,140){\line(1,0){25}}
\put(35,161){\line(1,0){10}}
\put(45,140){\line(0,1){21}}
\put(36,163){\scriptsize $\lambda_0^{(1)}\lambda_0^{(2)}$}
\put(55,140){\line(0,1){9}}
\put(45,149){\line(1,0){10}}
\put(46,151){\scriptsize $\lambda_1^{(1)}\lambda_1^{(2)}$}

\put(45,128){+}

\put(10,110){$(ii)$}
\put(35,90){\line(0,1){27}}
\put(35,90){\line(1,0){25}}
\put(35,108){\line(1,0){10}}
\put(36,110){\scriptsize $\lambda_0^{(1)}\lambda_1^{(2)}$}
\put(45,90){\line(0,1){18}}
\put(55,90){\line(0,1){12}}
\put(45,102){\line(1,0){10}}
\put(46,104){\scriptsize $\lambda_1^{(1)}\lambda_0^{(2)}$}

\put(45,80){\vector(0,-1){10}}

\put(10,30){$(iii)$} 
\put(35,10){\line(0,1){45}}
\put(35,10){\line(1,0){25}}
\multiput(35.25,28)(1.5,0){7}{\circle*{.3}}
\put(45,10){\line(0,1){39}}
\put(55,10){\line(0,1){21}}
\multiput(45.25,22)(1.5,0){7}{\circle*{.3}}
\put(35,49){\line(1,0){10}}
\put(38,51){\scriptsize $\lambda_0^{(1)}$}
\put(45,31){\line(1,0){10}}
\put(48,33){\scriptsize $\lambda_1^{(1)}$}

\put(45,3){(a)}


\put(85,140){\line(0,1){27}}
\put(85,140){\line(1,0){25}}
\put(85,161){\line(1,0){10}}
\put(95,161){\line(0,-1){12}}
\put(105,140){\line(0,1){9}}
\put(85,149){\line(1,0){20}}

\put(95,128){+}

\put(85,90){\line(0,1){27}}
\put(85,90){\line(1,0){25}}
\put(85,108){\line(1,0){10}}
\put(95,108){\line(0,-1){6}}
\put(105,90){\line(0,1){12}}
\put(85,102){\line(1,0){20}}

\put(95,80){\vector(0,-1){10}}
 
\put(85,10){\line(0,1){45}}
\put(85,10){\line(1,0){25}}
\put(95,49){\line(0,-1){18}}
\put(105,10){\line(0,1){21}}
\put(85,49){\line(1,0){10}}
\put(85,31){\line(1,0){20}}

\put(95,3){(b)}

\end{picture}

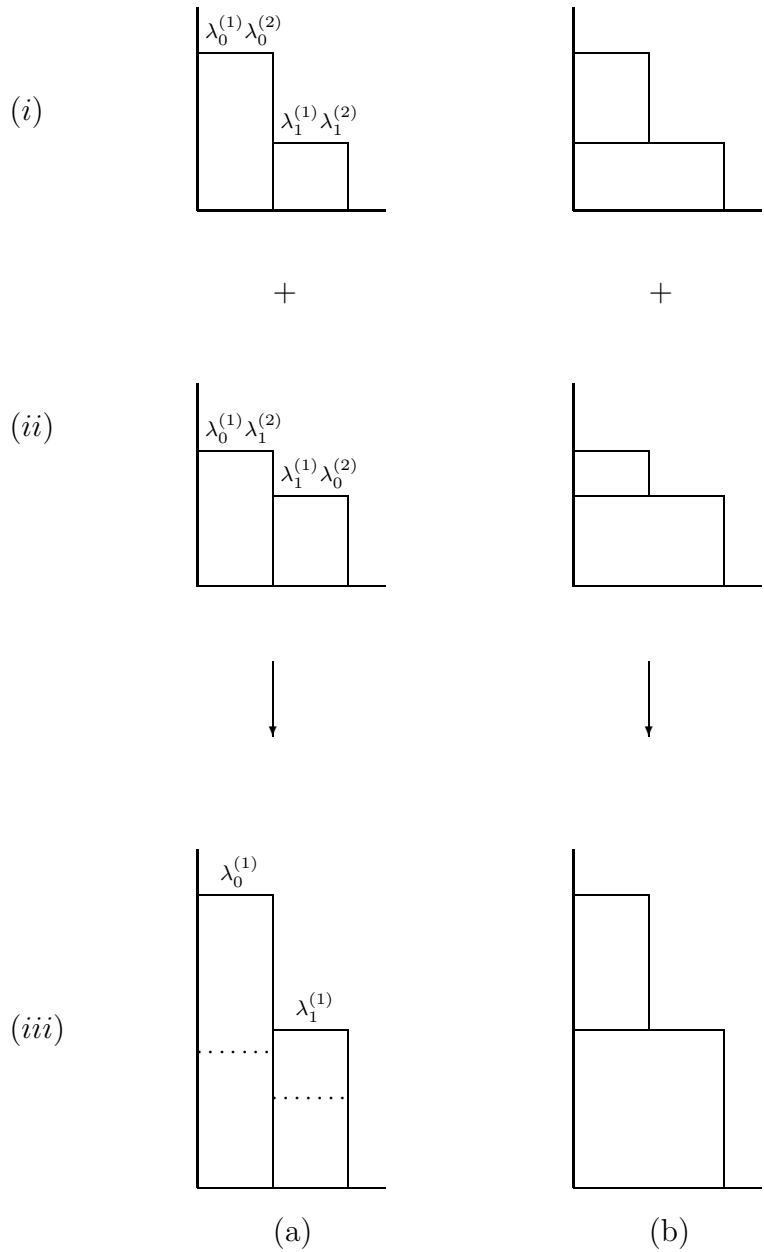
Figure 2: The area diagrams for two 2-level states are shown. The columns have unit width and 
the number above each column indicates the height.

\section{$m\;  \times\;  m$ case}
In this section, we generalise the $2 \times 2$ case considered in the
previous section to the $m \times m$ case, i.e. an entangled state
between two $m$-level systems.  Let us consider two entangled pairs of particles (1,2) and (3,4).
 Assume 1 and 4 are spatially separated as in Figure 1.  $m$-level entangled states for (1,2) and (3,4) are
\bea
|\psi^{(1)}\RR_{12} &=& \sum_{i=0}^{m-1} \sqrt{\lambda_i^{(1)}} |ii\RR_{12}  \label{m12}\\
|\psi^{(2)}\RR_{34} &=& \sum_{i=0}^{m-1} \sqrt{\lambda_i^{(2)}} |ii\RR_{34}
\label{m34}\eea
We can assume that the $\lambda$'s are ordered such that
$\lambda_i^{(n)} \geq \lambda_{i+1}^{(n)}$  for $n=1,2$ without loss of
generality.  
Next, particles 2 and 3 are brought together and generalised Bell measurements are performed. 
We generalise the Bell basis as follows  
\be
|\psi_{\alpha}^{\beta} \RR_{23}= \frac{1}{\sqrt m} \sum_{l=0}^{m-1}  e^{((l\beta) {\rm mod} \, m )2\pi i /m } |l,l+\alpha\RR_{23}
\ee
where we abbreviated $|(l + \alpha ) {\rm mod}\,  m\RR$ to $|l+\alpha\RR$ 
(which we will assume throughout this paper).
For $\alpha, \beta = 0, \cdots, m-1$, 
it can be checked that $\LL \psi_{\alpha^{\prime}}^{\beta^{\prime}} |\psi^{\beta}_{\alpha}\RR = \delta_{\alpha^{\prime} \alpha} \delta^{\beta^{\prime} \beta}$,
 thereby yielding $m^2$ orthonormal states.
With this basis, $|\psi^{(1)}\RR_{12}\otimes |\psi^{(2)}\RR_{34}$ is projected onto
\be
|\psi_{proj}\RR =\frac{1}{\sqrt{P_{\alpha}}} \frac{1}{\sqrt m}  \sum_{l=0}^{m-1}  e^{-((l\beta) {\rm mod} \, m )2\pi i/m } \sqrt{ \lambda_l^{(1)} \lambda_{l+\alpha}^{(2)}} |l, l+\alpha\RR_{14}
\ee
with probability $P_{\alpha}$, which can be obtained from normalisation.
Note that ${\rm mod}\, m$ arithmetic is assumed for subscripts of $\lambda$
  (which we will assume throughout the remainder of this paper).
The (unnormalised) Schmidt coefficients
for each $\alpha=0,\cdots,  m-1$ are as follows ( $\beta$ does not make any difference since the phase term disappears),
\be
\lambda_0^{(1)} \lambda_{0+\alpha}^{(2)}\;\;\; \; ,\;\;\; \;  \lambda_1^{(1)} \lambda_{1+\alpha}^{(2)}\;\;\; , \cdots, \;\;\; \lambda_{m-1}^{(1)} \lambda_{m-1+\alpha}^{(2)}
\label{lambda}\ee
Now we want to re-order (\ref{lambda}) such that 
the highest value  is set equal to $Z_0^{\alpha}$, and the next highest
is set equal to $Z_1^{\alpha}$,
and so on until the lowest which is set equal to $Z_{m-1}^{\alpha}$.
 Then the average entanglement obtained from the formula (\ref{Emax}) is given as 
\be
E^{\rm max} = \sum_{\alpha=0}^{m-1} \sum_{i=1}^m (Z_{i-1}^{\alpha} - Z_i^{\alpha} ) i \log_2 i
\label{ZZmm}\ee
where $Z_m^{\alpha}=0$.
Note that the probabilities $P_{\alpha}$ have cancelled with the
normalisation constants in this formula.

Let us consider an example with $m=3$.  From (\ref{m12}) and (\ref{m34}), we have the following states,
\bea
|\phi\RR_{12} & = & \sqrt{\lambda_0^{(1)}} |00\RR_{12} + \sqrt{\lambda_1^{(1)}} |11\RR_{12} + \sqrt{\lambda_2^{(1)}} |22\RR_{12} \\
|\phi\RR_{34} & = & \sqrt{\lambda_0^{(2)}} |00\RR_{34} + \sqrt{\lambda_1^{(2)}} |11\RR_{34} + \sqrt{\lambda_2^{(2)}} |22\RR_{34} 
\eea
A generalised Bell measurement is made on particles 2 and 3.
The unnormalised Schmidt coefficients given in (\ref{lambda})
will be obtained with $m=3$ and $\alpha=0,1,2$.  Suppose we impose the
following additional condition on the terms
in (\ref{lambda}), 
\bea
(i) \; \; \lambda_0^{(1)}\lambda_0^{(2)} & \geq & \lambda_1^{(1)} \lambda_1^{(2)}  \geq  \lambda_2^{(1)} \lambda_2^{(2)} \nonumber \\
(ii) \; \; \lambda_0^{(1)}\lambda_1^{(2)} & \geq & \lambda_1^{(1)} \lambda_2^{(2)} \geq  \lambda_2^{(1)} \lambda_0^{(2)} \label{ii} \\
(iii) \; \; \lambda_0^{(1)}\lambda_2^{(2)} & \geq & \lambda_1^{(1)} \lambda_0^{(2)}  \geq  \lambda_2^{(1)} \lambda_1^{(2)} \nonumber
\eea
where $(i),(ii)$ and $(iii)$ correspond to $\alpha=0,1,2$ respectively.
This means that there is no re-ordering when the $Z$'s are assigned.
The Schmidt coefficients in $(i),(ii),(iii)$ can be represented on graphs as shown in (a) of Figure 3. 
 $(i)$ is always true since coefficients of $|\phi\RR_{12}$
 and $|\phi\RR_{34}$ are ordered.  However $(ii)$ and $(iii)$ may not always be the case.  
Nevertheless, if we assume
$(i), (ii)$ and $(iii)$ to hold, then as in the 2 state case, we can add the boxes of $(i), (ii)$ and $(iii)$ 
and obtain $(iv)$ in 
Figure 3 and calculate the maximum average entanglement for obtaining a maximal state from $(iv)$ which is just 
$\sum_{j=1}^3 (\lambda_{j-1}^{(1)} - \lambda_j^{(1)} ) j \log_2 j$, the $E^{\rm max}$ from $|\phi\RR_{12}$.

%
%
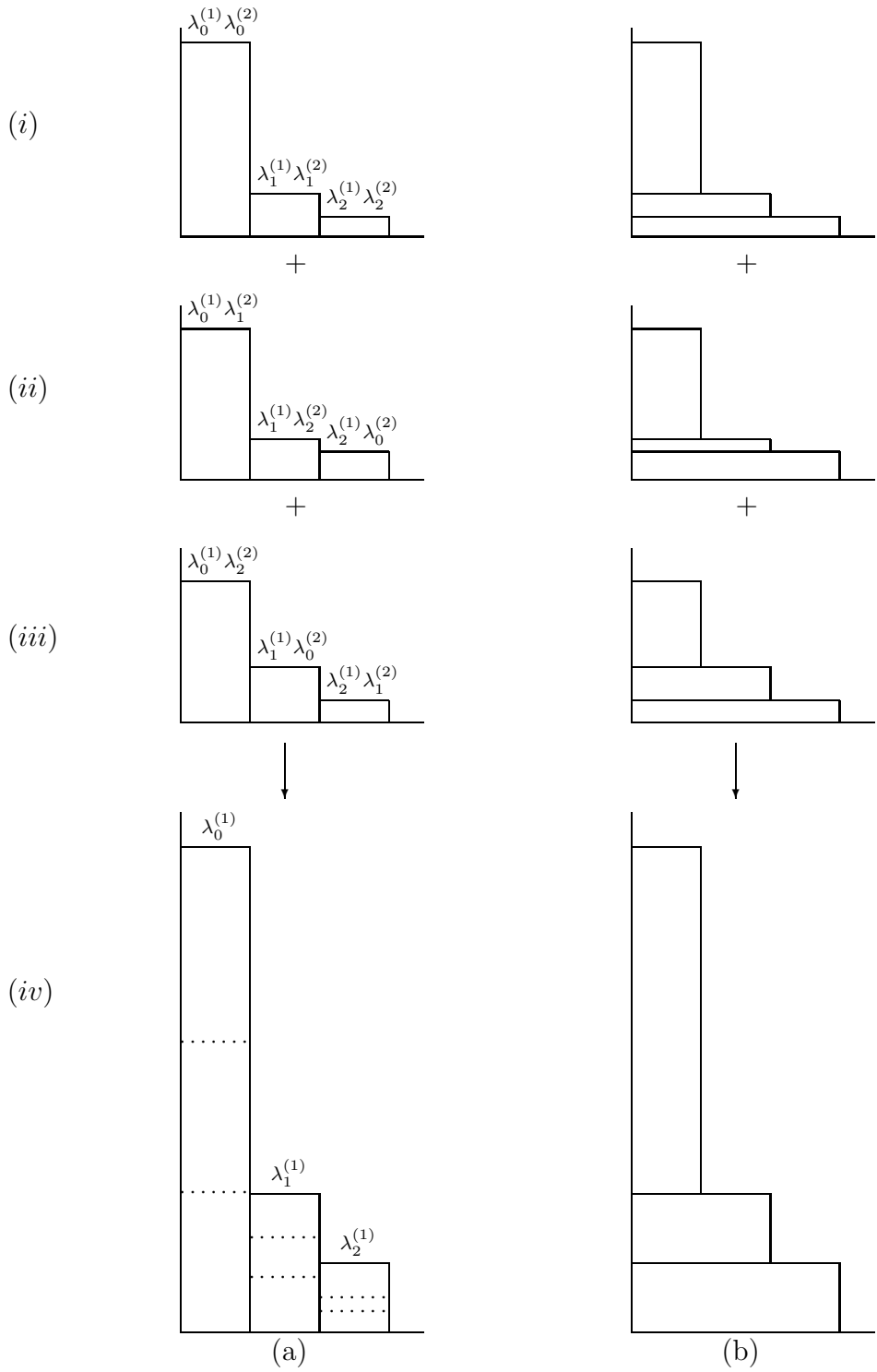
\begin{picture}(120,200)

\put(5,55){$(iv)$}
\put(30,7){\line(1,0){35}}
\put(30,7){\line(0,1){75}}
\put(40,7){\line(0,1){70}}
\multiput(30.25,27.3)(1.5,0){7}{\circle*{.3}}
\multiput(30.25,49)(1.5,0){7}{\circle*{.3}}
\put(30,77){\line(1,0){10}}
\put(33,79){\scriptsize $\lambda_0^{(1)}$}

\put(50,7){\line(0,1){20}}
\multiput(40.25,15)(1.5,0){7}{\circle*{.3}}
\multiput(40.25,20.8)(1.5,0){7}{\circle*{.3}}
\put(40,27){\line(1,0){10}}
\put(43,29){\scriptsize $\lambda_1^{(1)}$}

\put(60,7){\line(0,1){10}}
\multiput(50.25,10.1)(1.5,0){7}{\circle*{.3}}
\multiput(50.25,12.1)(1.5,0){7}{\circle*{.3}}
\put(50,17){\line(1,0){10}}
\put(53,19){\scriptsize $\lambda_2^{(1)}$}

\put(45,92){\vector(0,-1){8}}

\put(5,106){$(iii)$}
\put(30,95){\line(1,0){35}}
\put(30,95){\line(0,1){25}}
\put(40,95){\line(0,1){20.3}}
\put(30,115.3){\line(1,0){10}}
\put(31,117.3){\scriptsize $\lambda_0^{(1)}\lambda_2^{(2)}$}
\put(50,95){\line(0,1){8}}
\put(40,103){\line(1,0){10}}
\put(41,105){\scriptsize $\lambda_1^{(1)}\lambda_0^{(2)}$}
\put(60,95){\line(0,1){3.1}}
\put(50,98.1){\line(1,0){10}}
\put(51,100.1){\scriptsize $\lambda_2^{(1)}\lambda_1^{(2)}$}

\put(45,125){+}

\put(5,142){$(ii)$}
\put(30,130){\line(1,0){35}}
\put(30,130){\line(0,1){25}}
\put(40,130){\line(0,1){21.7}}
\put(30,151.7){\line(1,0){10}}
\put(31,153.7){\scriptsize $\lambda_0^{(1)}\lambda_1^{(2)}$}
\put(50,130){\line(0,1){5.8}}
\put(40,135.8){\line(1,0){10}}
\put(41,137.8){\scriptsize $\lambda_1^{(1)}\lambda_2^{(2)}$}
\put(60,130){\line(0,1){4}}
\put(50,134){\line(1,0){10}}
\put(51,136){\scriptsize $\lambda_2^{(1)}\lambda_0^{(2)}$}

\put(45,160){+}

\put(5,180){$(i)$}
\put(30,165){\line(1,0){35}}
\put(30,165){\line(0,1){30}}
\put(40,165){\line(0,1){28}}
\put(30,193){\line(1,0){10}}
\put(31,195){\scriptsize $\lambda_0^{(1)}\lambda_0^{(2)}$}
\put(50,165){\line(0,1){6.2}}
\put(40,171.2){\line(1,0){10}}
\put(41,173.2){\scriptsize $\lambda_1^{(1)}\lambda_1^{(2)}$}
\put(60,165){\line(0,1){2.9}}
\put(50,167.9){\line(1,0){10}}
\put(51,169.9){\scriptsize $\lambda_2^{(1)}\lambda_2^{(2)}$}

\put(43,3){(a)}

\put(95,7){\line(1,0){35}}
\put(95,7){\line(0,1){75}}
\put(105,77){\line(0,-1){50}}
\put(95,77){\line(1,0){10}}

\put(115,27){\line(0,-1){10}}
\put(95,27){\line(1,0){20}}

\put(125,7){\line(0,1){10}}
\put(95,17){\line(1,0){30}}

\put(110,92){\vector(0,-1){8}}

\put(95,95){\line(1,0){35}}
\put(95,95){\line(0,1){25}}
\put(105,115.3){\line(0,-1){12.3}}
\put(95,115.3){\line(1,0){10}}
\put(115,103){\line(0,-1){4.9}}
\put(95,103){\line(1,0){20}}
\put(125,95){\line(0,1){3.1}}
\put(95,98.1){\line(1,0){30}}

\put(110,125){+}

\put(95,130){\line(1,0){35}}
\put(95,130){\line(0,1){25}}
\put(105,151.7){\line(0,-1){15.9}}
\put(95,151.7){\line(1,0){10}}
\put(115,135.8){\line(0,-1){1.8}}
\put(95,135.8){\line(1,0){20}}
\put(125,130){\line(0,1){4}}
\put(95,134){\line(1,0){30}}

\put(110,160){+}

\put(95,165){\line(1,0){35}}
\put(95,165){\line(0,1){30}}
\put(105,193){\line(0,-1){21.8}}
\put(95,193){\line(1,0){10}}
\put(115,171.2){\line(0,-1){3.3}}
\put(95,171.2){\line(1,0){20}}
\put(125,165){\line(0,1){2.9}}
\put(95,167.9){\line(1,0){30}}

\put(108,3){(b)}

\end{picture}

Figure 3: The area diagrams for two 3-level states are shown.  The columns have unit width and 
 the number above each column indicates the height.

\vspace{1cm}

Since the $\lambda$'s are assumed to be ordered initially, it follows that the columns in (\ref{ii}) are
 already ordered 
(e.g. $\lambda_0^{(1)} \lambda_0^{(2)} \geq \lambda_0^{(1)} \lambda_1^{(2)}$) and 
decrease from top to bottom.
 Hence, inequalities can be compressed into 
\be
\lambda_{i_1}^{(1)} \lambda_{i_2}^{(2)} \geq \lambda_{i_1^{\prime}}^{(1)} \lambda_{i_2^{\prime}}^{(2)}
\label{conditionMM}\ee
if and only if
\be
i_1 i_2 \leq i_1^{\prime} i_2^{\prime}
\label{conditionMM1}\ee
where $i_1 i_2$ is interpreted as a number in base 3 (i.e. it is equal to $3^1 i_1 + 3^0 i_2$).
In fact, this can be done for any $m$ where $i_1 i_2$ is
 now interpreted as a number in base $m$ (i.e. equal to $m^1 i_1 + m^0 i_2$).
 Then $E^{\rm max}$ can be obtained from $|\phi\RR_{12}$ only, 
i.e. $\sum_{j=1}^m (\lambda_{j-1}^{(1)} - \lambda_j^{(1)}) j \log_2 j$.  
What are the class of $\lambda$'s which would 
satisfy the condition in (\ref{conditionMM}) and (\ref{conditionMM1})?  
One trivial example is when $|\psi^{(2)}\RR$ is maximally entangled state, so that all $\lambda^{(2)}$'s are 
$\frac{1}{m}$.
In the next section, we generalise the above procedure to the $N$-chain of 
$m$-level entangled states and we give nontrivial class of cases which satisfy the condition in (\ref{conditionMM}) and (\ref{conditionMM1}).

In the cases where (\ref{conditionMM}) and (\ref{conditionMM1}) are satisfied we have again
 that the result is optimal and that we can obtain maximal entanglement 
equal to that extractable from the weakest link in the chain.  
However, when (\ref{conditionMM}) and (\ref{conditionMM1}) do not hold then we do not know that (\ref{ZZmm}) is 
optimal since it is possible that a different measurement by Bob could yield better results.

%
%

\section{$N$-chained case}
In this section, we generalise the two $m\times m$ entangled states
to $N$-chained $m\times m$ states.
 As shown in Fig. 4, there are $N$ such states and the
measurements are made on particles $(2,3),(3,4),(5,6),\cdots,(2N-2,2N-1)$, so that
we are left with a single entangled pair between the particles 1 and $2N$.  We would like to know 
what the highest average entanglement that can be concentrated in the form of maximally entangled states
between $1$ and $2N$ is.

\begin{picture}(120,95)

\put(5,55){\framebox(137,30)}
\put(14,72){\line(1,0){20}}
\put(14,72){\circle*{2}}
\put(13,67){1}
\put(34,72){\circle*{2}}
\put(33,67){2}
\put(39,72){\line(1,0){20}}
\put(39,72){\circle*{2}}
\put(38,67){3}
\put(59,72){\circle*{2}}
\put(58,67){4}
\put(64,72){\line(1,0){20}}
\put(64,72){\circle*{2}}
\put(63,67){5}
\put(84,72){\circle*{2}}
\put(83,67){6}
\put(89,72){\circle*{1}}
\put(92,72){\circle*{1}}
\put(95,72){\circle*{1}}
\put(98,72){\circle*{1}}
\put(101,72){\circle*{1}}
\put(104,72){\circle*{1}}
\put(112,72){\line(1,0){20}}
\put(112,72){\circle*{2}}
\put(106,67){$2N-1$}
\put(132,72){\circle*{2}}
\put(129,67){$2N$}


\put(72.5,52.5){\vector(0,-1){10}}

\put(5,10){\framebox(137,30)}
\put(14,27){\line(1,0){118}}
\put(14,27){\circle*{2}}
\put(13,22){1}
\put(132,27){\circle*{2}}
\put(129,22){$2N$}

\end{picture}

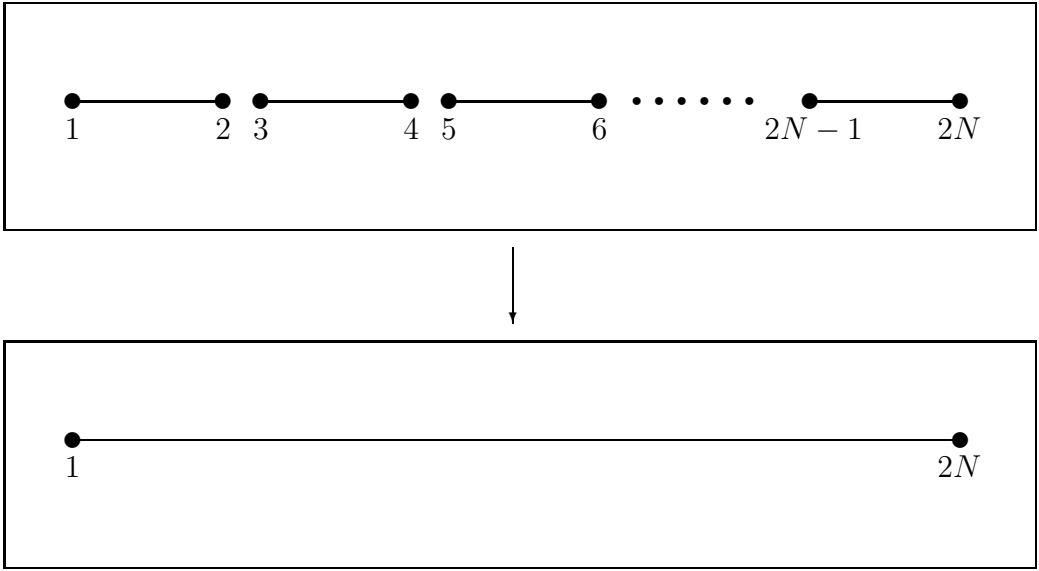
Figure 4:   The swapping of entanglement of $N$-chained states is shown.  Generalised Bell measurements 
are performed at each intermediate site $(2,3),(4,5),\cdots , (2N-2,2N-1)$.
\vspace{1cm}
\newline
We start with the following  $m\times m$ entangled states,
\bea
|\varphi^{(1)}\RR &=& \sum_{i=0}^{m-1} \sqrt{\lambda_i^{(1)}} |ii\RR_{12} \nonumber \\
|\varphi^{(2)}\RR &=& \sum_{i=0}^{m-1} \sqrt{\lambda_i^{(2)}} |ii\RR_{34} \label{varphiN} \\
                        &\vdots & \nn \\
|\varphi^{(N)}\RR &=& \sum_{i=0}^{m-1} \sqrt{\lambda_i^{(N)}} |ii\RR_{2N-1,2N} \nn
\label{chainstates}\eea  
There are $N-1$ intermediate locations which we label by $n=1$ to $N-1$.
At each location, a measurement is made onto the generalised Bell
basis $|\psi_{\alpha_n}^{\beta_n}\RR_{2n,2n+1}$.
If we define $\gamma_n \equiv \sum_{i=1}^n \alpha_i$, then
 $|\psi^{(1)}\RR \otimes \cdots \otimes |\psi^{(N)}\RR$
is projected as follows,
\be
|\psi_{proj}\RR=\frac{1}{\sqrt{P_{\alpha}}} \frac{1}{\sqrt m} \sum_{l=0}^{m-1}  e^{-((l\sum \beta_n ) {\rm mod}\, m )2\pi i /m } \sqrt{\lambda_l^{(1)}\lambda_{l+\gamma_1}^{(2)} \cdots \lambda_{l+\gamma_{N-1}}^{(N)}}|l,l+\gamma_{N-1}\RR_{1,2N}
\ee
with probability $P_{ \{\alpha_\} }$ which can be obtained from normalisation.
The unnormalised Schmidt coefficients for each outcome $\{ \alpha_i \}$ are,
\be
\lambda_0^{(1)} \cdots \lambda_{\gamma_{N-1}}^{(N)}\;\;\;\;  , \;\;\;\; \; \lambda_1^{(1)}  \cdots \lambda_{1+\gamma_{N-1}}^{(N)}\;\;\;\; , \;\;\;\;  \cdots \;\;\;\; , \;\;\;\; \lambda_{m-1}^{(1)} \cdots \lambda_{m-1+\gamma_{N-1}}^{(N)}
\label{lambda2}\ee
(again $\beta$ does not contribute since the phase terms disappear).
As in the previous section, 
the elements in (\ref{lambda2})  can be re-ordered (again as $Z_0\geq \cdots \geq Z_{N-1}$) and $E^{\rm max}$
can be calculated using the formula in (\ref{Emax}) for outcome $\{
\alpha_i \}$ then adding all of them up which gives
\be
E^{\rm max} = \sum_{\gamma_1=0}^{m-1} \cdots \sum_{\gamma_{N-1}}^{m-1} \sum_{i=1}^m \left( Z_{i-1}^{\gamma_1 \cdots \gamma_{N-1}} - Z_i^{\gamma_1 \cdots \gamma_{N-1}} \right)  i \log_2 i
\ee
between $1$ and $2N$.  This formula has the curious property that $E^{\rm max}$
is independent of the particular order that the states
(\ref{chainstates}) are in (since the $\gamma$'s in (\ref{lambda2}) take
all values).

Let us consider a special case such that,
\be
\lambda_{i_1}^{(1)} \lambda_{i_2}^{(2)} \cdots \lambda_{i_N}^{(N)} \geq \lambda_{i_1^{\prime}}^{(1)}  \lambda_{i_2^{\prime}}^{(2)} \cdots \lambda_{i_N^{\prime}}^{(N)}
\label{condition3}\ee
if and only if
\be
i_1 i_2 \cdots i_N \leq i_1^{\prime} i_2^{\prime} \cdots i_N^{\prime}
\label{condition33}\ee
where $i_1 \cdots i_N$ is an integer in base $m$ (i.e. $i_1 m^{N-1} + i_2 m^{N-2}+\cdots + i_N m^0$ ).
Then, by employing similar graphical reasoning to that in the
previous section, $E^{\rm max}$ can be obtained from $|\varphi^{(1)}\RR$ only,
i.e. 
\be
E^{\rm max}= \sum_{i=1}^m (\lambda_{i-1}^{(1)} - \lambda_i^{(1)} ) i \log_2 i
\ee
In such cases we obtain maximal entanglement extractable from the weakest link.
In this case, the weakest link is $|\varphi^{(1)}\RR$.  However since
$E^{\rm max}$ is independent of the order of the states,
similar results would apply in the case where
 one of the other links is the weakest.  In general the condition would be 
\be
\lambda_{i_1}^{(1)} \lambda_{i_2}^{(2)} \cdots \lambda_{i_N}^{(N)} \geq \lambda_{i_1^{\prime}}^{(1)}  \lambda_{i_2^{\prime}}^{(2)} \cdots \lambda_{i_N^{\prime}}^{(N)}
\label{condition4}\ee
if and only if 
\be
{\rm Perm}(i_1i_2\cdots i_N) \leq {\rm Perm}(i_1^{\prime} i_2^{\prime} \cdots i_N^{\prime} )
\label{condition44}\ee
where Perm permutes the positions of the entries in the base $m$ number.  The weakest link is then that 
corresponding to the leftmost entry in the string.

In fact, not only is the average entanglement concentrated equal to
average entanglement that can be concentrated from the weakest link, but
also the distribution of $|\varphi_m\RR$ states are the same.  This
follows directly from the area diagrams.

We will now show that there exists a nontrivial class of $\lambda$'s
which satisfy the condition in (\ref{condition3},\ref{condition33}).
Suppose $\Lambda$'s are unnormalised $\lambda$'s, i.e.
 $\lambda_j^{(l)} = \frac{\Lambda_j^{(l)}}{\left( \sum_{j=0}^{m-1}\Lambda_j^{(l)} \right)}$.
With the normalisation factor 
$L=\left( (\sum_{i_1=0}^{m-1} \Lambda_{i_1}^{(1)} ) \cdots ( \sum_{i_N=0}^{m-1} \Lambda_N^{i_N} )\right) $,
 the condition in (\ref{condition3}) and (\ref{condition33}) becomes  
\be
\frac{\Lambda_{i_1}^{(1)} \cdots \Lambda_{i_N}^{(N)}}{L} \geq \frac{\Lambda_{i_1^{\prime}}^{(1)} \cdots \Lambda_{i_N^{\prime}}^{(N)}}{L}
\label{aL}\ee
if and only if 
\be
i_1 \cdots i_N \leq i_1^{\prime} \cdots i_N^{\prime}
\label{Is}\ee
Taking logarithm with base $b > 1$, of (\ref{aL}), gives
\be
 \log_b \Lambda_{i_1}^{(1)}+\cdots + \log_b \Lambda_{i_N}^{(N)} \geq \log_b \Lambda_{i_1^{\prime}}^{(1)} + \cdots \log_b \Lambda_{i_N^{\prime}}^{(N)} 
\label{LOGlambda}\ee
(since $\log_b$ has positive gradient everywhere when $b>1$).
If we define $\bi \equiv (m-1) - i$ then (\ref{Is}) is equivalent to $\bi_1 \bi_2 \cdots \bi_N \ge \bi_1^{\prime}\bi_2^{\prime} \cdots \bi_N^{\prime}$.
Let us take a set 
of non-negative constants $\eta_n$ where $n=1,\cdots, N$ satisfying
$\eta_n \geq (m-1) \eta_{n+1}$. Then (\ref{Is}) implies
\be
\bi_1 \eta_1 + \cdots + \bi_N \eta_N \geq \bi_1^{\prime} \eta_1 + \cdots + \bi_N^{\prime} \eta_N
\label{NEWeta}\ee
Comparing (\ref{LOGlambda}) and (\ref{NEWeta}), we can put $\log_b \Lambda_{i_n}^{(n)} =\bi_n  \eta_n $. 
Then 
\be
\Lambda_{i_n}^{(n)} = b^{\bi_n  \eta_n}
\label{ConLam1}\ee
The $\lambda$'s can then be found by normalisation.
One example is where $\eta_{n-1} \equiv m \eta_n$ and $\eta_N =1$, then
\be
\Lambda_{i_n}^{(n)} = b^{\bi_n m^{N-n}}
\label{ConLam2}\ee
In this case the $\lambda$'s decrease more steeply for smaller $n$ and
less steeply for larger $n$ and hence $E^{\rm max}$ increases with $n$ for the
links in the chain.  Another interesting example is when $\eta_1=1$ and
$\eta_n=0$ for $n=2$ to $N$.  In this case the first entangled state in
the chain will be some non-maximally entangled state while the remaining
entangled states will all be maximally entangled (since their
$\Lambda$'s are equal to 1).  This is the well known situation in which
the entanglement of the first state is successively teleported along the
chain so that it is finally shared by the two end parties.

In those cases where (\ref{condition4},\ref{condition44}) is satisfied
we know that this method is optimal since no concentration
protocol could yield better results than that corresponding to the
weakest link.  However, in those cases where this condition is not
satisfied, we do not know that the protocol discussed here is optimal.
It is possible that other protocols in which measurements other than
generalised Bell measurements are made at the intermediate stages may
yield better results.

\section{Obtaining GHZ states}
So far we have only considered Bell measurements.  In this section, we
give a simple example that converts
two general $2\times 2$ bipartite states into a single GHZ state
\cite{GHZ} with some probability.
Let Alice and Bob share entangled particles 1 and 2 and Alice and Charlie
share 3 and 4 with the following states,
\bea
|\varphi^{(1)}\RR_{12}& = & \sqrt{\lambda_0^{(1)}} |00\RR_{12} + \sqrt{\lambda_1^{(1)}}|11\RR_{12} \nn \\
|\varphi^{(2)}\RR_{34}&  & \sqrt{\lambda_0^{(2)}}|00\RR_{34} + \sqrt{\lambda_1^{(2)}}|11\RR_{34}
\eea
where we take $\lambda_0^{(n)} \geq \lambda_0^{(n)}$ for $n=1,2$ and we assume
that $\lambda_0^{(1)}\geq\lambda_1^{(2)}$.
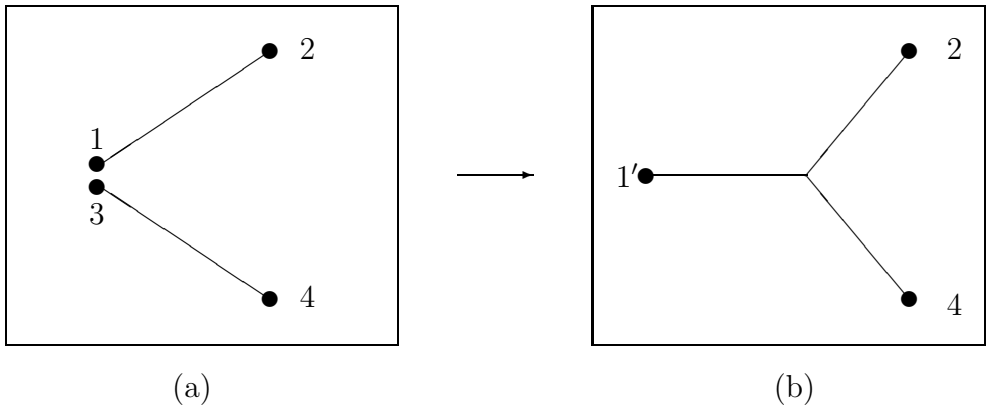
\setlength{\unitlength}{1mm}

\begin{picture}(120,80) 

\put(30,3){(a)}
\put(110,3){(b)}

\put(43,49){\line(-3,-2){22}}
\put(20,34){\circle*{2}}
\put(43,49){\circle*{2}}
\put(43,16){\circle*{2}}
\put(20,31){\circle*{2}}
\put(43,16){\line(-3,2){22}}
\put(19,36){1}
\put(19,26){3}
\put(47,48){2}
\put(47,15){4}
\put(8,10){\framebox(52,45)}

\put(68,32.5){\vector(1,0){10}}

\put(93,32.5){\line(1,0){21.5}}
\put(93,32.5){\circle*{2}}
\put(128,49){\line(-5,-6){13.8}}
\put(128,16){\line(-5,6){13.8}}
\put(128,49){\circle*{2}}
\put(128,16){\circle*{2}}
\put(89,31){$1^{\prime}$}
\put(133,48){2}
\put(133,14){4}
\put(86,10){\framebox(52,45)}

\end{picture}

Figure 5: The conversion of two bipartite states to a single GHZ state is shown. A degenerate measurement is  
made on particles 1 and 3.
 
\vspace{1cm}
Alice makes measurements on her particles 1 and 3 of the following
(degenerate) operators
\bea
F_1 &=& |00\RR_{13} \LL 00| + |11\RR_{13}\LL 11| \nn \\
F_2 &=& |01\RR_{13} \LL 01| + |10\RR_{13} \LL 10| 
\eea
Then with probability  $P_1 = \lambda_0^{(1)}\lambda_0^{(2)} + \lambda_1^{(1)}\lambda_1^{(2)}$ 
Alice, Bob and Charlie are left with 
\be
\frac{1}{\sqrt{P_1}}(\sqrt{\lambda_0^{(1)}\lambda_0^{(2)}}|0000\RR_{1324} +
\sqrt{\lambda_1^{(1)}\lambda_1^{(2)}}|1111\RR_{1324} )
\ee
We can redefine the states$|00\RR_{13}$ and $|11\RR_{13}$ of
particles 1 and 3 which are in Alice's hands to
be $|0\RR_{1'}$ and $|1\rangle_{1'}$ respectively of system $1'$ (system
$1'$ consisting of particles 1 and 3). Then the state they share is
\be
\frac{1}{\sqrt{P_1}}(\sqrt{\lambda_0^{(1)}\lambda_0^{(2)}}|000\RR_{1'24} +
\sqrt{\lambda_1^{(1)}\lambda_1^{(2)}}|111\RR_{1'24} )
\ee
Also with probability $P_2 =\lambda_0^{(1)}\lambda_1^{(2)} +
\lambda_1^{(1)}\lambda_0^{(2)}$, the following state is obtained:
\be
\frac{1}{\sqrt{P_2}} (\sqrt{\lambda_0^{(1)}\lambda_1^{(2)}}|0101\RR_{1324}
+ \sqrt{\lambda_1^{(1)}\lambda_0^{(2)}}|1010\RR_{1324})
\ee
Again we can redefine $|01\RR_{13}\equiv |2\RR_{1'}$ and
$|10\RR_{13}\equiv |3\RR_{1'}$.
so the state is
\be
\frac{1}{\sqrt{P_2}} (\sqrt{\lambda_0^{(1)}\lambda_1^{(2)}}|201\RR_{1324}
+ \sqrt{\lambda_1^{(1)}\lambda_0^{(2)}}|310\RR_{1324})
\ee
It is possible to obtain a GHZ state from either of these two states by
performing a local filtering operation. The probability of this being
successful is equal to the twice the square of the smaller coefficient
in the expansion.
Since $\sqrt{\lambda_0^{(1)}\lambda_0^{(2)}}\geq \sqrt{\lambda_1^{(1)}\lambda_1^{(2)}}$
 and $\sqrt{\lambda_0^{(1)}\lambda_1^{(2)}} \geq \sqrt{\lambda_1^{(1)}\lambda_0^{(2)}}$,
the probability of obtaining a GHZ state is
\bea
&\Rightarrow & 2P_1 \frac{\lambda_1^{(1)}\lambda_1^{(2)}}{P_1} + 2P_2 \frac{\lambda_1^{(1)}\lambda_0^{(2)}}{P_2} \nn \\
&\Rightarrow & 2\lambda_1^{(1)}
\label{2b2}\eea
To see that this is optimal suppose we could do better. That is suppose the
parties can obtain a GHZ with higher probability than $2\lambda_1^{(1)}$.
Once they have this GHZ Charlie could make a measurement on his particle
which collapses the other two particles into $2\times 2$ dimensional
maximally entangled state thus obtaining such a state with probability
higher than $2\lambda_1^{(1)}$ but this contradicts equation (1) when
applied to Alice and Bob.


\section{Conclusions}

In this paper we have generalised entanglement swapping to a chain of
arbitrary pure entangled states in arbitrary dimensions with
generalised Bell measurements.  For a chain
consisting of two $2\times 2$ dimensional entangled states it has been
shown that the entanglement concentrated between the two ends is equal
to that concentratable from the weakest link in the chain.  For
longer chains and/or for higher dimensions this result does not hold.
However, there does exist a broad class of cases in which the
entanglement concentrated is equal to that concentratable from the
weakest link.  This proves that, in these cases, entanglement swapping
(with generalised Bell measurements) is an optimal way of concentrating
entanglement. However, we do not know that it is optimal in general.
Nevertheless it interesting is that there is a much richer
structure when we go to higher dimensions and longer chains.

In this paper we have addressed the matter of entanglement manipulation
by LOCC in a particular setting.  Namely where we have chains of pure bipartite
entangled states and we wish to concentrate to maximally entangled
states.  The most general setting would be where a number of parties
share a general entangled state and want to manipulate it by LOCC to
some other state or distribution of states.  However, there are many
restricted situations (such as a chain of bipartite pure
states as considered here) that fall short of this most general
setting. By considering other restricted cases we may hope to gain a
deeper understanding general entanglement manipulation.
The example in section 5 can be considered as work in
this direction.

\vspace{2cm}

{\bf Acknowledgements}.  LH is funded by a Royal Society University
Research Fellowship.

\vfill
\newpage

\end{document}